# Synthesis of single-layered fluorographdiyne nanosheets via selective on-surface 2D covalent polymerization

Chen-Hui Shu[1,#,*], Yi Zheng[5,#], Tao Lin[6,*], Li-Xia Kang[3], Zhang Qu, Zhi-Yu Wang, Ying Wang[3], Zheng-Yang Huang[3], Qian Liu[3], Hang Xu[1], Chong Chen[1], Yangfan Wu[4,5], Longteng Xiao[4,5], Mengxi Liu[4,5,*], Xiaohui Qiu[4,5], Pei-Nian Liu[2,3,*], and Deng-Yuan Li[2,*]

[1]School of Future Technology, Henan University, 475004 Kaifeng, China.

[2]State Key Laboratory of Natural Medicines, School of Pharmacy, China Pharmaceutical University, 211198 Nanjing, China.

[3]School of Chemistry and Molecular Engineering, East China University of Science & Technology, 200237 Shanghai, China.

[4]Laboratory of Standardization and Measurement for Nanotechnology, National Center for Nanoscience and Technology,100049 Beijing, China.

[5]University of Chinese Academy of Sciences, 100049 Beijing, China.

[6]College of New Materials and New Energies Shenzhen Technology University, 518118 Shenzhen, China.

[#]These authors contribute equally to this work.

*Corresponding Authors: ch_shu@foxmail.com, lintao@sztu.edu.cn, liumx@nanoctr.cn, liupn@cpu.edu.cn, dengyuanli@cpu.edu.cn.

## Table of Contents artwork

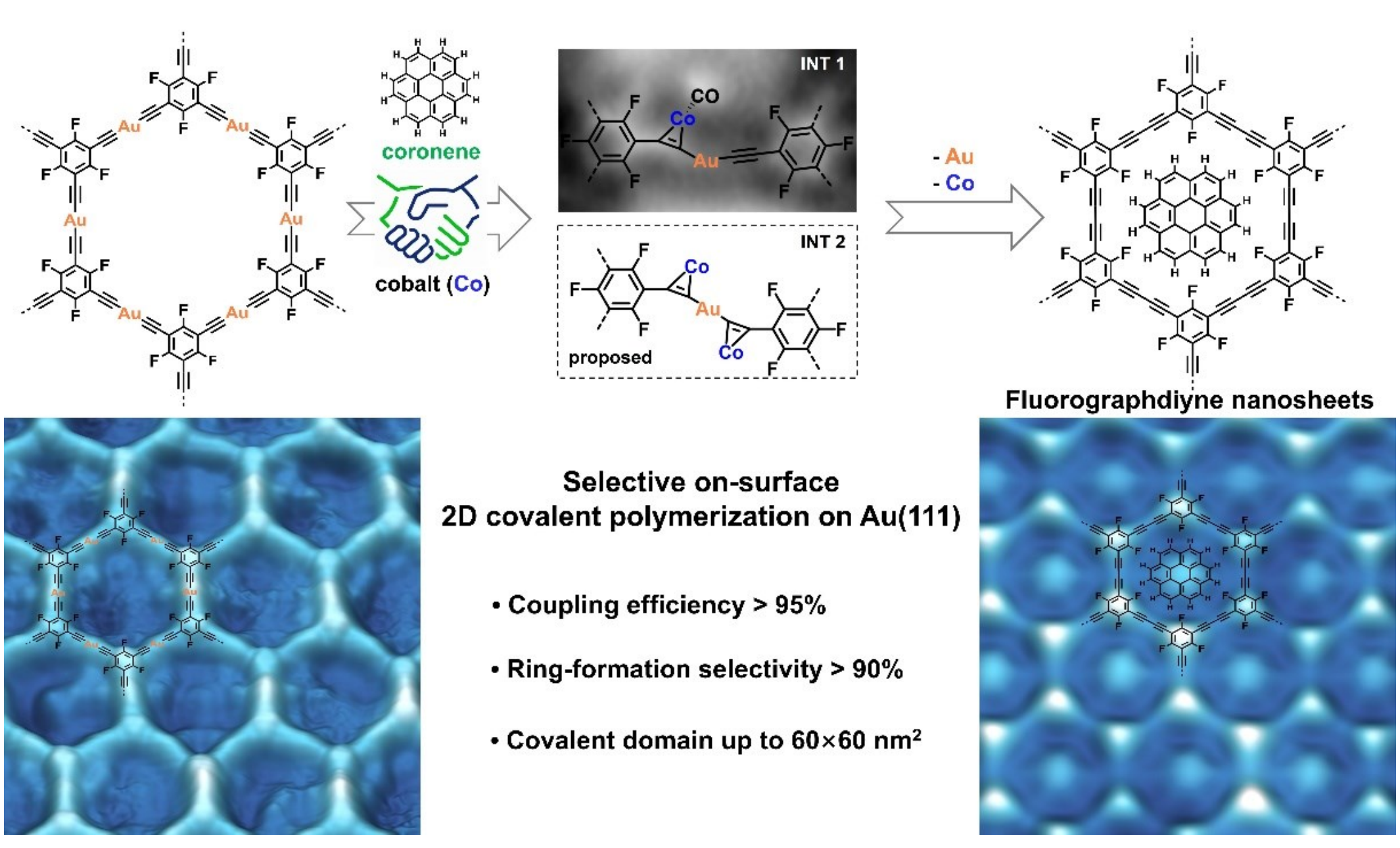

## Abstract

Two-dimensional conjugated polymers (2DCPs) are significant macromolecular materials with intriguing and tunable physicochemical properties that depend on their geometries. Graphdiyne and its derivatives are exemplary 2DCPs featuring $sp$-$sp^2$ hybridized skeletons. However, achieving single-layered, large-domain/regular graphdiyne and its derivatives on surfaces remains a formidable challenge due to the lack of selective 2D covalent polymerization methods. Here, we report a selective on-surface 2D covalent polymerization method via the combination of cobalt catalysis and coronene templating, achieving the synthesis of single-layered fluorographdiyne nanosheets up to 60×60 nm$^2$ on Au(111) surface. Using scanning probe techniques, we visualize the sequential polymerization process and characterize cobalt-activated coupling intermediates at the atomic level. Experimental and theoretical analyses suggest that strong d-π coupling between cobalt and alkynyl transforms a robust $C_{sp}$-Au bond into a weaker $C_{sp2}$-Au bond, thereby facilitating the demetallization C-C coupling. Besides, the templating effect of coronene suppresses kinetically trapped defects and improves the selectivity of hexagonal-ring formation in the complex 2D covalent polymerization process.

## Introduction

Two-dimensional conjugated polymers (2DCPs)[1-3], a class of crystalline layered covalently-linked macromolecular sheets, have gained rapid recognition in many cutting-edge fields[2-7] owing to their well-defined geometries and tunable physicochemical properties. In recent years, on-surface 2D covalent polymerization[8] has received considerable attention not only for the construction of distinctive single-layered 2DCPs with fascinating properties such as excellent conductance[9], tunable bandgaps (metallic to semiconducting)[10,11], and Dirac cones or flat bands[12-14], but also for facilitating an in-depth understanding of growth mechanisms—a critical prerequisite for achieving their large-domain selective synthesis and future applications. Generally, achieving selective on-surface 2D covalent polymerization requires the precise regulation of thermally-driven processes including molecular assembly, the activation of functional groups or intermediates, diffusion, the formation of covalent bonds and rings. To address that, strategies such as hierarchical constructions[10,15-18], thermodynamic controls[19-21] and the use of customized precursors[9] have been developed, yielding remarkable results in the synthesis of novel single-layered 2DCPs[10,14-15,22-30]. Despite these advances, however, the large-domain selective on-surface synthesis of single-layered 2DCPs so far still defines one of the major challenges in surface chemistry, as it requires simultaneous improvement of coupling efficiency, coupling selectivity, and ring-formation selectivity in complex 2D covalent polymerization involving a large number of precursors.

Owing to the $p_\pi$-$d_\pi$ overlap between alkynyl and metal atom, which can effectively stabilize potential polymerization intermediates[31], large-domain *sp*-hybridized metal-organic networks (*sp*-MONs) offer a perfect example to investigate the polymerization process and mechanism of constructing 2DCPs on surfaces. In addition, the high carrier transport of carbon-carbon triple bonds[32-34] and the rich topologies of *sp*-*sp*$^2$ hybridized systems render alkyne-bridged 2DCPs[35-37] important 2D carbon-based materials. However, the high thermostability of large-domain *sp*-MONs and the complex reactions of alkynyl[24,38] pose a major challenge for the fabrication of well-ordered alkyne-bridged 2DCPs on surfaces. Furthermore, the flexible nature of alkyne motifs will inevitably lead to the formation of kinetically-trapped ring defects upon annealing[30]. These defects deform the polymerization skeletons and disrupt subsequent covalent-ring formation, thereby inducing more structural defects. In fact, such negative effects will continuously accumulate during large-domain synthesis in the absence of effective regulation, ultimately leading to uncontrollable polymerization behavior. Efforts have been devoted to steering 2D covalent polymerization for fabricating various alkyne-bridged 2DCPs on surfaces, including the use of templates[16,24], the halogen/oxygen-assisted promotion[25,39], and functional group transformations[26,40]. To date, however, achieving efficient demetallization coupling and selective ring formation of large *sp*-MONs has rarely been explored, which leaves the large-domain synthesis of alkyne-bridged 2DCPs on surfaces as a severe challenge.

Exogenous transition metals have been shown to enhance the efficiency and selectivity of various important on-surface reactions[41-43], while thus far their application in facilitating on-surface 2D covalent polymerization remains largely unexplored. In this study, cobalt (Co) was employed to induce the transformation of robust $C_{sp}$-Au bond to weak $C_{sp2}$-Au bond via the strong d-π orbital coupling between Co and alkynyl (Fig. 1a)[44,45], thus achieving the efficient demetallization C-C coupling of large-domain *sp*-MONs. Besides, coronene was used to direct the ring-formation process, achieving high selectivity of hexagonal ring formation (Fig. 1b). By steering 2D covalent polymerization with a synergistic strategy combining Co catalysis and coronene templating, we synthesized the coronene-embedded fluorographdiyne nanosheets with domain sizes up to 60×60 $nm^2$ on Au(111) surface from *sp*-MONs (Fig. 1c). Additionally, by the comprehensive investigations of scanning tunneling microscopy (STM), non-contact atomic force microscopy (nc-AFM), and theoretical analyses, we visualized the covalent polymerization process at atomic level, and elucidated the underlying mechanisms for the efficient Co-catalyzed demetallization C-C coupling and the coronene-assisted ring-formation regulation.

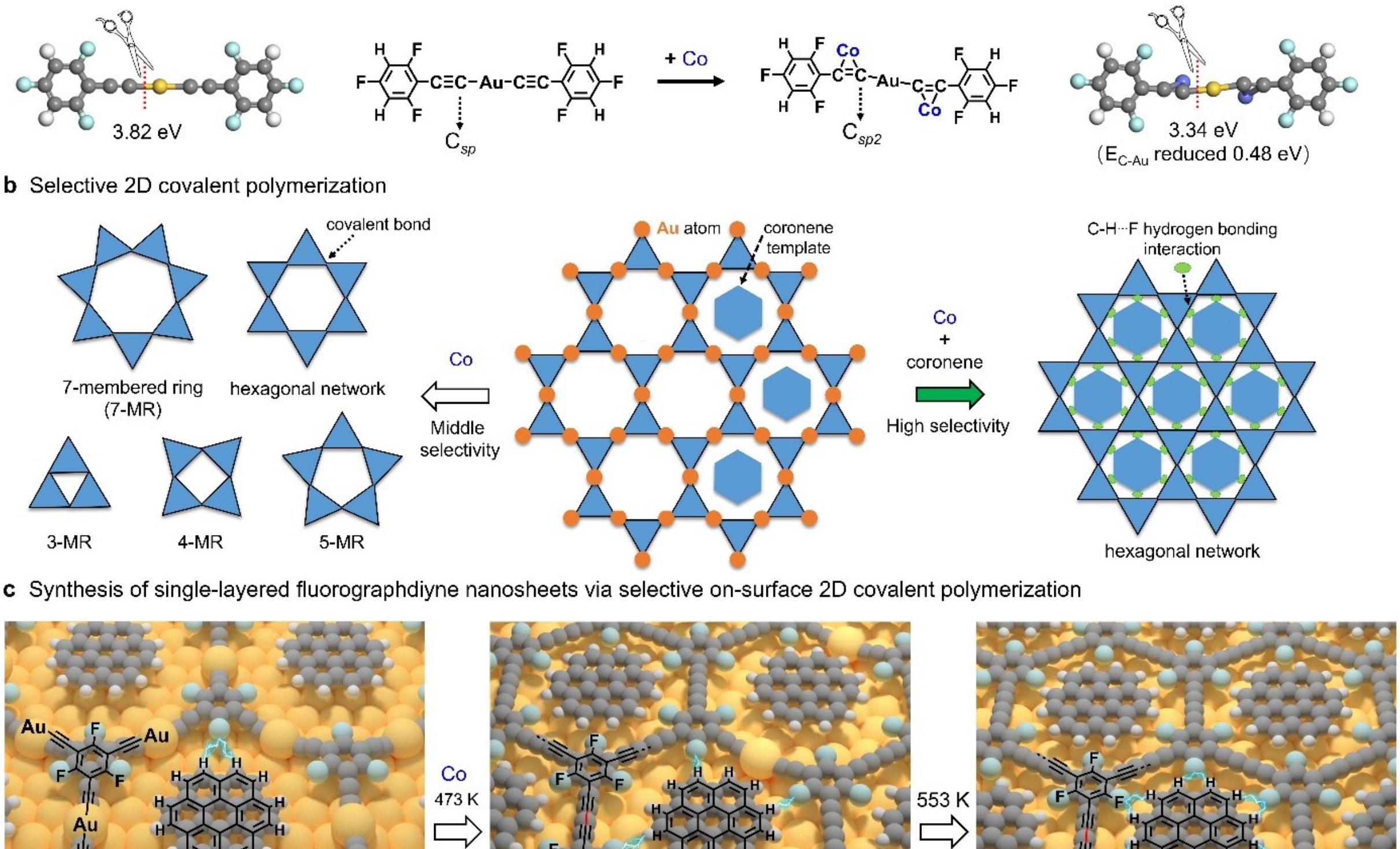

**Fig. 1 | Synthesis of single-layered fluorographdiyne nanosheets via selective 2D covalent polymerization on Au(111)**. **a**, Weakening the alkynyl-Au bond by the coordination interaction between Co and alkynyl. **b**, Selective 2D covalent polymerization via combining Co catalysis and coronene templating. **c**, Synthesis of single-layered fluorographdiyne nanosheets via selective on-surface 2D covalent polymerization.

## Results and Discussion

### Cobalt-catalyzed polymerization of *sp*-MONs

After depositing 0.7 monolayer (ML) 1,3,5-tris(chloroethynyl)-2,4,6-trifluorobenzene (tFtCEB) on Au(111) surface held at 250 K, a flower-like self-assembled structure composed of alkynyl-Au-alkynyl dimer formed (Fig. 2a and 2e). Owing to the strong electron-withdrawing effect of fluorine which improves the reversibility of the $C_{sp}$-Au bond, a large-domain honeycomb *sp*-MONs formed after annealing sample to 473 K (Supplementary Fig. 1). This result is in good consistency with the low barrier of 0.62 eV to break a $C_{sp}$-Au bond (Supplementary Fig. 2). Co (0.05 ML) was deposited onto *sp*-MONs, and the sample was subsequently annealed to 473 K. As depicted in Fig. 2b, many bright islands formed in the network and the atomic image of an island was inserted. According to the measured lattice constant of 0.343 nm[46] (Supplementary Fig. 3) and the disappeared Cl atoms in *sp*-MONs (Fig. 2f), these bright islands are assigned to $CoCl_2$ islands. Further adding 0.05 ML Co and annealing the sample to 473 K, an intermediate network containing ~50 % covalent linkages formed (Fig. 2c), suggesting that Co boosted the efficiency of covalent coupling. By coating alkynyl-Au-alkynyl linkages with gray, many deformed organometallic/covalent rings were highlighted. The white lines (covalent linkage) and gray dashed lines in Fig. 2g sketch a deformed network containing five-membered ring (5-MR) and concomitant organometallic 7/8-MR defects. After annealing sample to 553 K, a polycrystalline structure formed (Fig. 2d). nc-AFM characterization suggests the formation of fluorographdiyne at regular domains (Fig. 2h). Notably, statistical analysis shows that more than 95% of the alkynyl-Au-alkynyl linkages were transformed into the diacetylene linkages, illustrating the high coupling efficiency and coupling selectivity of the Co-catalyzed 2D covalent polymerization. However, the covalent polycrystalline structures also demonstrate that the large regular *sp*-MONs and Co catalysis are insufficient for achieving selective on-surface 2D covalent polymerization to construct large-domain fluorographdiyne nanosheets (Supplementary Fig. 4).

To understand the underlying catalysis mechanism of Co and capture the possible intermediates, we deposited 0.1 ML Co onto *sp*-MONs held at 250 K and then cooled the sample to 4 K for nc-AFM characterizations (Fig. 2i). Except the intact alkynyl-Au-alkynyl linkage (i-1), two possible Co-activated structures (i-2, i-4) were observed. According to previous reports[44,45], Co can bind to one π bond of alkynyl and form two C-Co bonds to transform $C_{sp}$ to $C_{sp2}$ configurations. Inspired by such information, our STM and AFM simulation analyses (Supplementary Fig. 5 and 6) suggest that the dashed circle of i-2 can be attributed to the complexes composed of Co-activated alkynyl and CO molecule (adsorbed onto the Co atom), respectively. For the disconnected structure (i-4), the identical feature marked by a dashed circle was also revealed in the nc-AFM image, suggesting the same activated configuration (Supplementary Fig. 6). In consideration of the weaker $C_{sp2}$-Au bond compared with $C_{sp}$-Au bond, we proposed that an i-3 intermediate may demetallize Au adatom and form i-4.

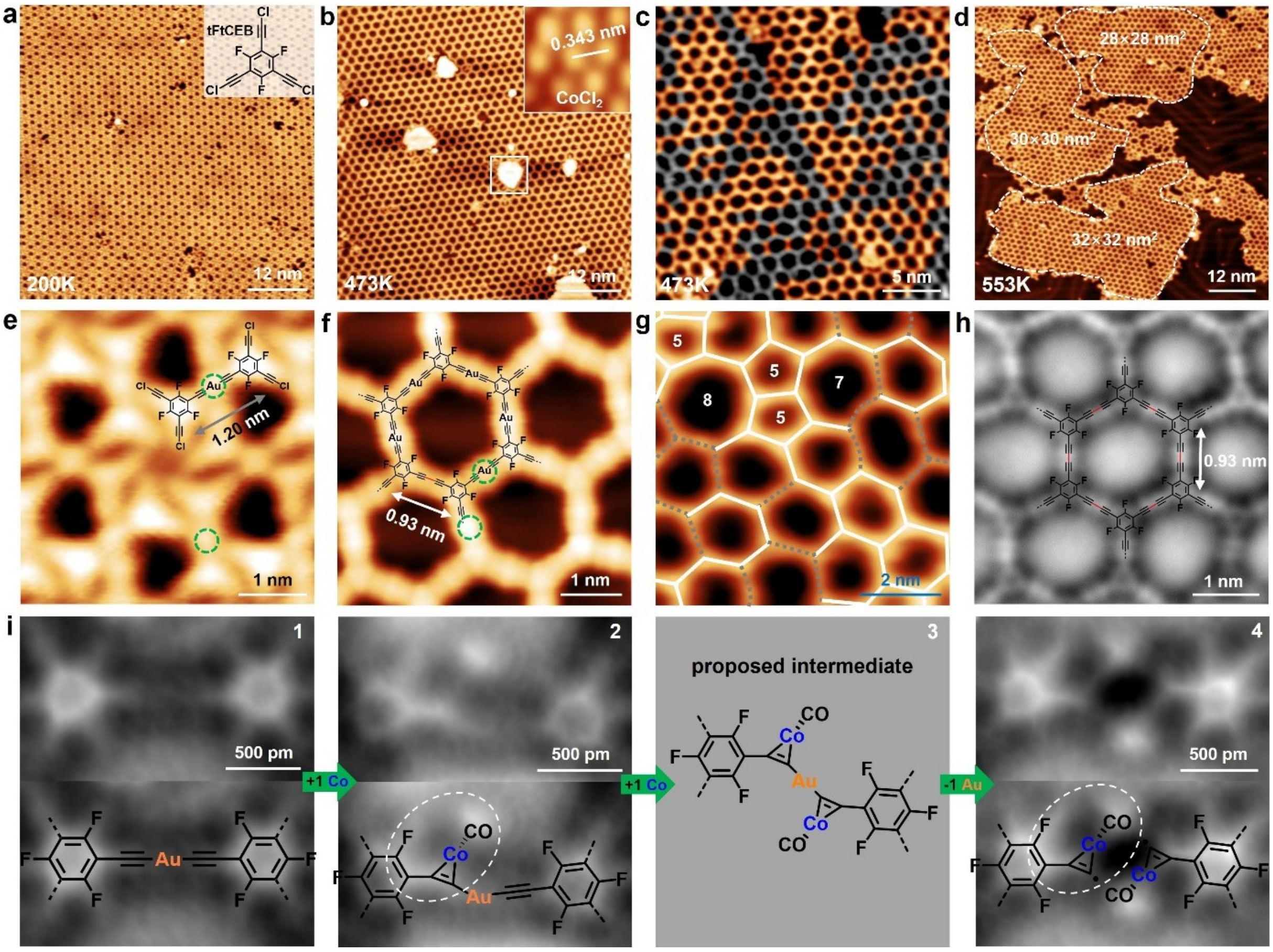

**Fig. 2 | Cobalt-catalyzed demetallization C-C coupling of *sp*-MONs on Au(111)**. **a**, **e**, Depositing tFtCEB on the Au(111) surface held at 250 K (a: -0.6 V, 100 pA; e: -0.4 V, 100 pA). **b**, **f**, Large-domain and zoom-in STM images after dosing 0.05 ML Co and annealing to 473 K (b: -1.5 V, 60 pA; f: -1.21 V, 150 pA). **c**, **g**, Large-domain and zoom-in STM images after adding 0.05 ML Co and annealing to 473 K (c: -1.5 V, 50 pA; c: -1.5 V, 20 pA). Orange and gray structures are covalently-coupled regions and alkynyl-Au-alkynyl regions, respectively. **d**, **h**, Large-scale STM and nc-AFM images after annealing to 553 K (d: -1.5 V, 30 pA). i) Observed intermediates involving Co-catalyzed alkynyl-Au-alkynyl activation.

## Selective 2D covalent polymerization via combing Co catalysis and coronene templating

As depicted in Fig. 3a, theoretical calculations demonstrate an excellent topological match between coronene and fluorographdiyne framework, and suggest the existence of multiple C–H···F hydrogen bonding interactions[47-49]. As shown in Fig. 3b and 3f, coronene was deposited onto large-domain *sp*-MONs. After this, 0.1 ML Co was deposited on coronene-embedded *sp*-MONs and then annealed the sample to 473 K. The partially-polymerized intermediate networks formed, containing ~50 % alkynyl-alkynyl coupling (Fig. 3c and 3g). Upon further annealing the sample to 553 K, large-domain coronene-embedded fluorographdiyne nanosheets formed (Fig. 3d). A high-resolution STM presents the uniform configuration of coronene in the fluorographdiyne lattice (Fig. 3h). The domain size of this ordered nanosheet is up to 60×60 $nm^2$, and other observed large nanosheets are provided in Supplementary Fig.7. Some defective locations are marked by blue dashed circles, and the ordered covalent skeletons adjoined these defective areas suggest that the disturbance of defects toward the polymerization was effectively suppressed by coronene's templating effect. Besides, black vacancies in the fluorographdiyne nanosheets can be attributed to the desorption of coronene during the high temperature annealing. Note that *sp*-MONs or coronene-embedded *sp*-MONs cannot form ordered fluorographdiyne nanosheets in the absence of Co catalysis (Supplementary Fig. 8).

Next, statistical analyses of ring-formation selectivity and the distribution of ordered domain sizes were employed to evaluate the quality of fluorographdiyne nanosheets constructed via the two synthetic methods described above. In the absence of coronene (red), the percentages of 6-MR and other non-hexagonal rings are 72.7% and 27.3%, respectively (Fig. 3e). In contrast, the blue histograms (with coronene) exhibit a high selectivity of 6-MR formation (92.3%) and low proportions (7.7%) for other non-hexagonal rings. As illustrated in Fig. 3e, to facilitate statistical analysis, the ordered domain sizes are divided into seven ranges: $8^2$-$15^2$ $nm^2$, $15^2$-$25^2$ $nm^2$, $25^2$-$35^2$ $nm^2$, $35^2$-$45^2$ $nm^2$, $45^2$-$55^2$ $nm^2$, $55^2$-$65^2$ $nm^2$, and $65^2$-$80^2$ $nm^2$. In the absence of coronene (red), the corresponding percentages were 46.1%, 34.8%, 14.6%, 3.7%, 0.8%, 0%, and 0%, respectively. When coronene was used (blue), the corresponding percentages were 19.5%, 28.6%, 27.9%, 13.4%, 7.0%, 2.4%, and 1.2%, respectively. The higher selectivity of 6-MR formation and larger ordered domains highlight the superiority of the Co/coronene synergistic strategy in achieving precise polymerization regulation for the selective synthesis of large-domain fluorographdiyne nanosheets on Au(111) surface. Additionally, we also examined the covalent polymerization of *sp*-MONs constructed from tCEB (without F) with the same regulation on Au(111) surface. Due to the absence of multiple C–H···F hydrogen bonding interactions and medium-quality *sp*-MONs, the quality of resulting coronene-embedded graphdiyne is slightly higher than that obtained with Co catalysis alone (Supplementary Fig. 9). Furthermore, the significant temperature difference of the coronene desorption experiments suggested the existence of multiple C–H···F hydrogen bonding interactions between fluorographdiyne and coronene (Supplementary Fig. 10).

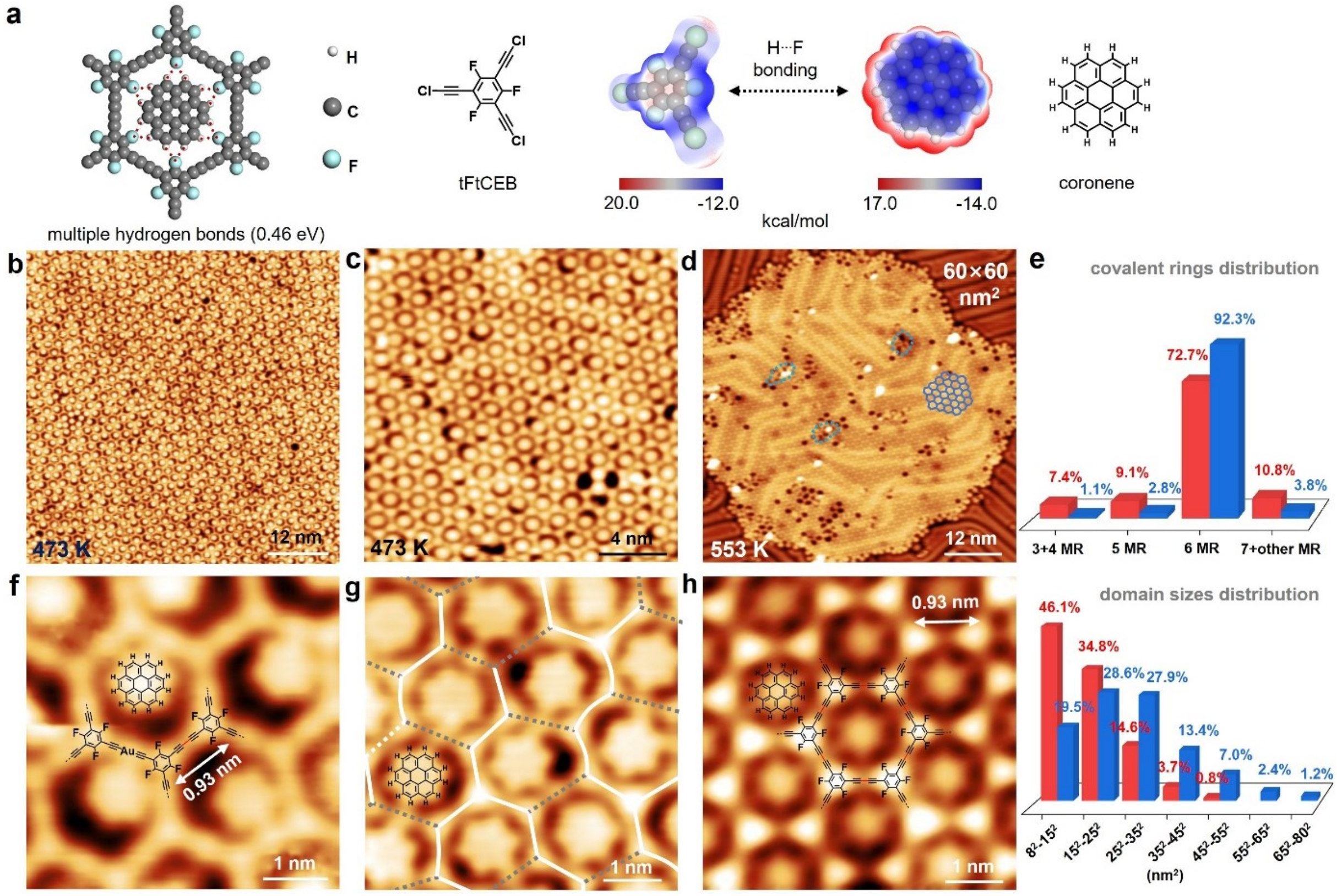

**Fig. 3 | Synthesizing fluorographdiyne nanosheets on Au(111) via selective 2D covalent polymerization based on the combination of Co catalysis and coronene templating**. **a**, The free-standing model of fluorographdiyne embedded coronene, and the molecular electrostatic potential (ESP) surface of used molecules. **b**, f, Large-scale and zoom-in images of coronene-embedded *sp*-MONs after annealing to 473 K (b: 1.0 V, 50 pA; f: 1.1 V, 50 pA). **c**, **g**, Large-scale and zoom-in images of coronene-embedded *sp*-MONs after dosing 0.1ML Co and annealing to 473 K (c: 1.5 V, 60 pA; g: -1.7 V, 150 pA). **d**, **h**, Large-scale and zoom-in images of fluorographdiyne nanosheet after annealing to 553 K (d: -1.3 V, 50 pA; h: -1.5 V, 300 pA). **e**, Statistical analysis of covalent rings and covalent domain sizes. Red for Co catalysis alone and blue for combing Co catalysis and coronene templating.

To characterize the electronic structures of coronene-embedded fluorographdiyne, differential conductance (dI/dV) spectra were performed on three representative positions (Fig. 4a and 4d). As depicted in Fig. 4b, two resonance states were identified at -0.90 V and 2.15 V at the diacetylene (green dot), benzene (blue) and the adjacent edge (purple) between coronene and fluorographdiyne. In addition to the aforementioned states, new resonance states were observed at -1.30 V and 2.40 V specifically at the purple dot position. We further performed the dI/dV mappings in a constant-height mode, facilitating a rational comparison with DFT-calculated local density of states (LDOS). The states at -0.90 V and 2.15 V are primarily localized on the fluorographdiyne frame, and their electronic features are consistent with the LDOS maps of VB (-0.23 V) and CB (2.90 V) of a periodic free-standing fluorographdiyne (Fig. 4c and 4e). This result indicates that the single-layered fluorographdiyne is a semiconductor with a bandgap of approximately 3.05 V. Furthermore, the state at -1.30 V is mainly distributed on the coronene moiety, presenting a six-ovals flower-like pattern. The state at 2.40 V exhibits a doughnut-like ring structure that merged with the fluorographdiyne frame (Fig. 4c). Moreover, the electronic features of -1.30 V and 2.40 V are analogous to the LDOS maps of the HOMO (-0.23 V) and LUMO (3.34 V) of a free-standing coronene (Fig. 4c and 4e), demonstrating a measured energy gap of approximately 3.70 V for the coronene.

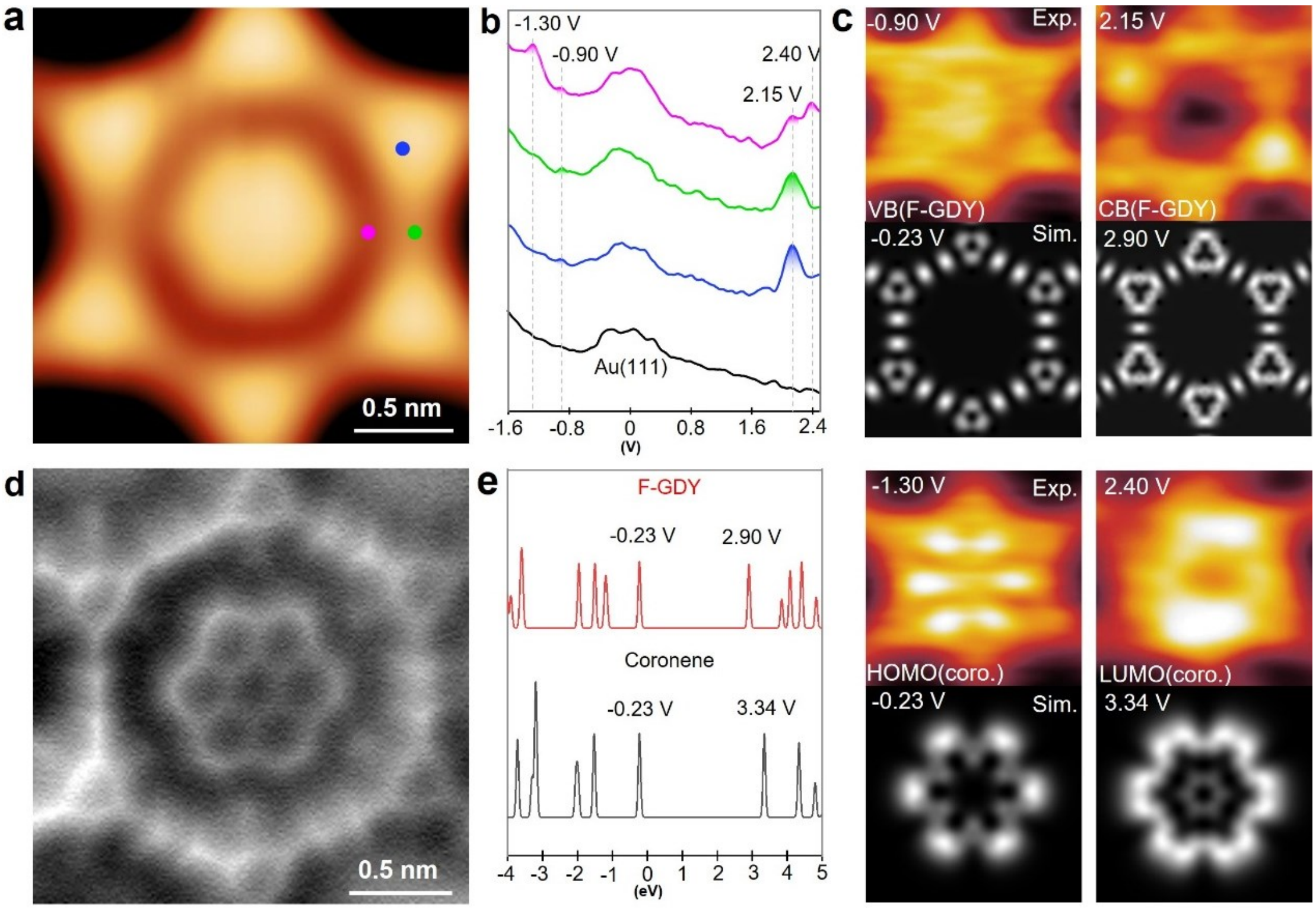

**Fig. 4 | Electronic structure of coronene-embedded fluorographdiyne on Au(111)**. **a**, **d**, The zoom-in STM and nc-AFM images of a coronene-embedded fluorographdiyne (a: -0.6 V, 20 pA). **b**, **e**, The dI/dV spectra recorded on different locations marked in the STM image, and the corresponding density of states of free-standing fluorographdiyne and coronene. **c**, Corresponding constant-height dI/dV maps conducted at energies marked in b, and calculated LDOS maps of LUMO and HOMO of free-standing fluorographdiyne (F-GDY) and coronene (coro.).

### Regulation mechanisms of cobalt and coronene in selective 2D covalent polymerization

To investigate the demetallization C-C coupling mechanism on Au(111), DFT calculations were performed using alkynyl-Au-alkynyl dimers as simplified models. In the absence of Co, the calculated barrier for direct demetallization coupling of the adatom-bound alkynyl-Au-alkynyl dimer is 2.63 eV (Fig. 5a), which is 0.59 eV greater than that of surface-stabilized alkynyl-Au-alkynyl dimer (Fig. 5b, 2.04 eV). These results suggest that $IS_1$ will dissociate Au adatom to form the surface-stabilized $IS_2$, which is supported by the highest barrier of 1.26 eV in the Au dissociation process (Supplementary Fig. 11). In the presence of Co, the possible reaction pathway for the demetallization coupling (Fig. 5c) was proposed based on previous studies[44,45] and corresponding bond-resolution nc-AFM characterizations. First, the surface-stabilized alkynyl-Au-alkynyl dimer ($IS_2$) undergoes coordination activation by two Co atoms, leading to the transformation of $C_{sp}$-Au bonds (C-C distance: 1.255 Å for $IS_2$ + 2Co) to $C_{sp2}$-Au bonds (C-C distance: 1.342 Å for $IS_3$). DFT calculations show that the coordination activation is an exothermic process with an energy release of -3.14 eV. Furthermore, the calculated coupling barrier of the coordination-activated $IS_3$ is only 1.27 eV, which is lower than that (2.04 eV) for the conversion of $IS_2$ to $FS_2$. These results further demonstrate that the coordination activation between alkynyl and Co atom weakens the $p_\pi$-$d_\pi$ orbital overlap of the alkynyl-Au bond, dramatically reduces the reaction barrier for the demetallization C-C coupling of large-domain *sp*-MONs on Au(111).

To understand the regulation mechanism of coronene in the covalent polymerization process, the evolution of the polymerization intermediates was characterized and analyzed by nc-AFM, DFT calculations, and atom-in-molecule (AIM) analysis[47-49]. Based on the number of coupling reactions in a six-membered ring, these intermediates and the final covalent product are labelled as C-0 to C-6. The top of Fig. 6a presents the successive coupling pathway from C-0 to C-6, in which the alkynyl-alkynyl couplings are marked with red dots. Coronene molecules are closely adsorbed at the rims of these intermediates, and green dashed lines mark the locations of potential noncovalent bonds. As depicted in the bottom of Fig. 6a, AIM analysis identifies noncovalent interactions in the corresponding models, such as C–H···F, C–H···Au, and C–H···π noncovalent interactions. The calculated total energies of these noncovalent interactions in the intermediates and final covalent product range from 0.19 eV to 0.46 eV.

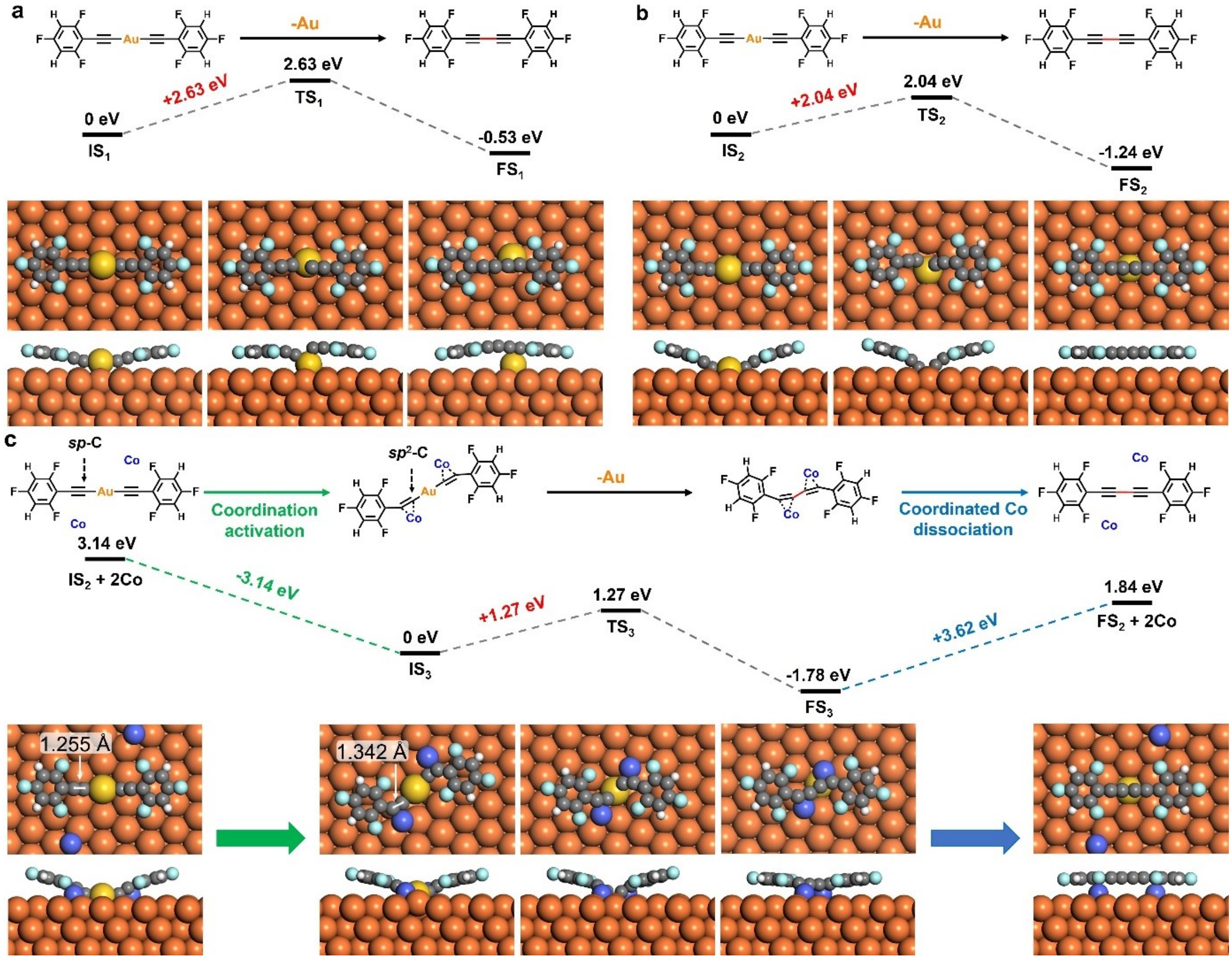

**Fig. 5 | DFT-calculated reaction pathways for the demetallization C-C couplings on Au(111)**. **a**, The coupling of adatom-bound alkynyl-Au-alkynyl dimer. **b**, The coupling of surface-stabilized alkynyl-Au-alkynyl dimer. **c**, Proposed Co-catalyzed demetallization C-C coupling of surface-stabilized alkynyl-Au-alkynyl dimer.

According to the results of nc-AFM characterizations and corresponding theory calculations, we proposed the possible mechanism of polymerization regulations on Au(111) (Fig. 6b). Under Co catalysis and the regulation of coronene, *sp*-MONs undergoes partial demetallization C-C couplings to reach a low-polymerized stage, in which activated oligomer and monomer radicals coexisted. Due to the occupation of embedded coronene, less space is available for the rotation/diffusion (indicated by dashed arrows) of radical species, thereby increasing the selectivity for hexagonal intermediates formation. In the absence of coronene, however, more space is available for the mobility of radical species, thus leading to a higher probability of forming kinetically-trapped non-hexagonal defects. In the high-polymerized stage, a higher annealing temperature is required to ensure the diffusion of larger oligomeric radical moieties, and embedded coronene similarly promotes the formation of hexagonal products. Besides, the noncovalent bonds between coronene and oligomers act as a "glue" that likely provide additional driving force to facilitate the growth of intact fluorographdiyne nanosheets.

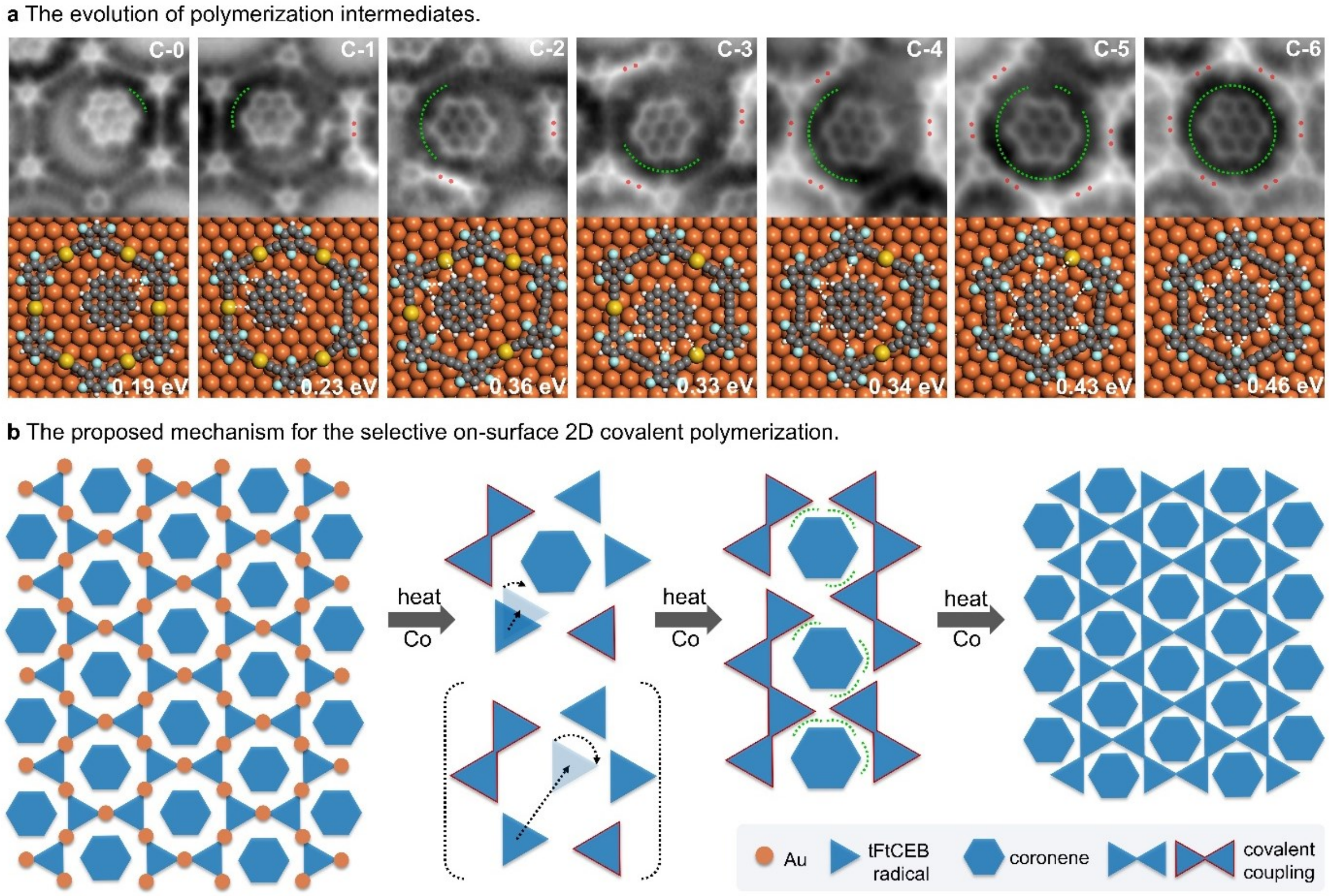

**Fig. 6 | Polymerization intermediates and proposed coronene's regulation of polymerization**. **a**, The evolution process of polymerization intermediates and corresponding theory calculations. **b**, Proposed mechanism for the selective on-surface 2D covalent polymerization.

## Conclusions

In summary, we have achieved the precise regulation of large-domain alkynyl-Au-alkynyl demetallization polymerization and synthesized ordered fluorographdiyne nanosheets with a size up to 60×60 nm$^2$ on Au(111) surface. Exogenous Co was chosen as a catalyst to activate and convert robust $C_{sp}$-Au bonds to weaker $C_{sp2}$-Au bonds through strong d-π orbital coupling between Co and alkynyl, thereby lowering the coupling barrier. Meanwhile, coronene was selected as a template to steer the covalent ring-formation process and enhance the selectivity of the covalent hexagonal ring. Through STM and nc-AFM investigations, we visualized the covalent polymerization process and characterized the polymerization intermediates at atomic level. Furthermore, combining the DFT-calculated reaction process and AIM analysis of the intermediates, we proposed a polymerization regulation mechanism of the synergistic method on Au(111) surface. This research not only provides a feasible reference for achieving precise control of on-surface 2D covalent polymerizations but also offers a promising approach for the functionalization and fabrication of innovative 2D materials.

## Methods

**Sample preparation and STM/AFM measurements.** Single crystalline Au(111) surface was cleaned by cycles of $Ar^+$ sputtering and annealing under UHV (base pressure $2\times10^{-10}$ mbar). tFtCEB and coronene molecules were synthesized and purchased from the commercial company. tFtCEB was evaporated from leaking valve (pre-heat to approximately 333 K) onto the Au(111) surface at room temperature. Coronene was evaporated from quartz crucible onto the Au(111) surface at room temperature, and the sublimation temperature was approximately 373 K. Then, the progression annealing was performed. nc-AFM, STS and partial STM measurements were performed on a ScientaOmicron UHV-LT-STM/nc-AFM system operated at ~5 K or Createc LT STM at ~77 K. STM images were taken in the constant-current mode using a tungsten tip. nc-AFM measurements were carried out in constant-height mode with a CO-functionalized tip (resonance frequency $f_0 \approx 40.7$ kHz, oscillation amplitude A ≈ 100 pm, quality factor $Q \approx 5.6 \times 10^4$). The dI/dV spectra were acquired by a lock-in amplifier, while the sample bias was modulated by a 553 Hz, 30 mV (r.m.s.) sinusoidal signal under open-feedback conditions. STM and nc-AFM images were processed by using Gwyddion software (http://gwyddion.net/download.php).

**Theoretical calculations.** The calculations were carried out in the framework of DFT by using the Vienna ab initio Simulation Package (VASP)[50]. The projector augmented-wave (PAW) method described the interaction between ions and electrons[51]. We used the generalized gradient approximation (GGA) with Perdew-Burke-Ernzerhof (PBE) formalism to treat exchange-correlation interaction[52], and van der Waals (vdW) interactions were considered by using the DFT-D3 developed by Grimme[53]. The energy cutoff for the plane wave basis sets is 400 eV, and the energy and force convergence value between two consecutive self-consistent steps were set as $10^{-5}$ eV and 0.05 eV/Å, respectively. The simulations of the reaction barriers in the manuscript were performed with the climbing image nudged elastic band (*CI*-NEB) method for finding saddle points and minimum energy paths, which is at T = 0 without including entropy and vibrations. The total coronene-frame binding energy ($E$) in the coronene-embedded intermediates or final covalent product is calculated as below[47],

$$E = E_{(frame+coronene+sub)} - E_{(sub+frame)} - E_{(sub+coronene)} + E_{sub}$$

in which $E_{(frame+coronene+sub)}$, $E_{(sub+frame)}$, $E_{(sub+coronene)}$, and $E_{sub}$ refer to the energies of the coronene-embedded intermediates or final covalent product, the frame on on the substrate, the coronene on the substrate, and the Au(111) substrate, respectively. Noncovalent bonds in the relaxed models were identified using AIM analysis[48] and *Multiwfn* software[49].

## Data availability

The authors declare that the main data supporting the findings of this study are available within the paper and its Supplementary Information files. Extra data are available from the corresponding authors upon request.

## Acknowledgements

This work was supported by the National Natural Science Foundation of China (Nos. 22272050, 92580139 and 22002044), the Shanghai Municipal Science and Technology Qi Ming Xing Project (No. 22QA1403000), and the Fundamental Research Funds for the Central Universities. This work was also supported by the National Key R&D Program of China (No. 2024YFA1208200), the National Natural Science Foundation of China (Nos. 22522201 and 22372048), and the CAS Project for Young Scientists in Basic Research (No. YSBR-054). We also thank Professor Weifeng Zhang for his help in STM characterization.

## Author contributions

C.-H.S., P.-N.L., and D.-Y.L. conceived and supervised the project; C.-H.S., H.X., C.C., Y.W., Z.-Y.H., and Q.L. performed on-surface synthesis experiments with the supervision of P.-N.L. and D.-Y. L.; Y.Z., Z.Q., Y.W., and L.X. performed nc-AFM and STS characterizations with the supervision of C.-H.S., T.L., M.L. and X.Q.; C.-H.S., T.L., L.-X.K., Z.Y.W and Y.W. conducted theoretical computations; C.-

H.S., M.L., and D.-Y.L. analyzed the data; C.-H.S. and D.-Y.L. wrote the manuscript; All authors discussed the results and helped write the manuscript at all stages.

## Competing interests

The authors declare no competing interests.

## Additional information

**Supplementary information.** Supplementary Information is linked to the online version of the paper at www.nature.com.

**Correspondence and requests for materials** should be addressed to Chen-Hui Shu, Tao Lin, Mengxi Liu, Pei-Nian Liu, and Deng-Yuan Li.